\documentclass[
    ,final            
  ]
  {aipproc}

\layoutstyle{6x9}

\begin{document}

\title{Diquark properties and their role in hadrons}

\author{P. Maris}{address={Dept. of Physics and Astronomy, 
University of Pittsburgh, Pittsburgh, PA 15260}}

\begin{abstract}
Diquark correlations are important in baryons, which can be modeled as
quark-diquark bound states.  In addition, diquarks could play a role
in non-standard hadrons such as tetraquarks and pentaquarks.  Here, we
obtain properties of these diquarks from the corresponding bound state
equation, using a model for the effective quark-quark interaction that
has proved successful in the light meson sector.  Subsequently, we use
the same model to estimate the masses of the lightest diquark-diquark
and diquark-antidiquark states.
\end{abstract}

\maketitle


%
The set of Dyson--Schwinger equations [DSEs] form a useful tool to
obtain a microscopic description of hadronic properties~\cite{review}.
The simplest hadrons are mesons: color-singlet bound states of a quark
and an antiquark.  They are described by solutions of the homogeneous
Bethe--Salpeter equation [BSE] for $q\bar{q}$ states.  The
Bethe--Salpeter amplitudes [BSAs] of different types of mesons, such
as pseudo-scalar, vector, etc. are characterized by different Dirac
structures.

In addition to $q\bar{q}$ bound states, one could also ask the
question whether or not there are $qq$ states, by studying the
corresponding BSE.  Two quarks can be coupled in either a color sextet
or a color antitriplet.  Single gluon exchange leads to an interaction
that is attractive for diquarks in a color antitriplet configuration.
Furthermore, it is the diquark in a color antitriplet that can couple
with a quark to form a color-singlet baryon.  Thus we only consider
$qq$ states in a color antitriplet configuration.  As in the case of
mesons, the different types of diquarks are characterized by different
Dirac structures.  Since the intrinsic parity of a $qq$ pair is
opposite to that of a $q\bar{q}$ pair, a scalar diquark BSA has
exactly the same form as a pseudoscalar meson BSA.

For practical calculations, we have to make a truncation of the set of
DSEs~\cite{review}.  Here we adopt the rainbow truncation of the quark
DSE, in combination with the ladder truncation for the BSE
\begin{eqnarray}
 \Gamma(p_+,p_-) &=& 
	- f_C \, \int_q^\Lambda\! {\cal G}(k^2) \; D_{\mu\nu}^{\rm free}(k)
	\gamma_\mu \, S(q_+) \, \Gamma_M(q_+,q_-) \, S(q_-) \, \gamma_\mu \; ,
\label{eq:ladderBSEm}
\end{eqnarray}
where $p_\pm = p \pm P/2$ and $q_\pm = q \pm P/2$ are the relative
quark momenta, $P$ the total meson momentum, $S(q_\pm)$ the
nonperturbatively dressed quark propagator, $D_{\mu\nu}^{\rm
free}(k=p-q)$ the free gluon propagator in Landau gauge and ${\cal
G}(k^2)$ a phenomenological effective
interaction~\cite{Maris:1997tm,Maris:1999nt}.  The color factor $f_C$
in the BSE is different for mesons and diquarks: $f_C = \frac{4}{3}$
for mesons, whereas $f_C = \frac{2}{3}$ for diquarks.

%
For the effective interaction, we use the model of
Ref.~\cite{Maris:1999nt}.  Our results for the light meson and diquark
masses are given in Table~\ref{tab:masses}.  The pseudoscalar and
vector meson properties seem to be rather independent of the details
of the effective interaction, as long as the interaction generates the
observed amount of chiral symmetry breaking~\cite{Maris:1999nt}.
However, the diquark masses are more sensitive to details of the
effective interaction.
\begin{table}[ht]
\begin{tabular}{ll|ccccc|cc|c}
\multicolumn{2}{l|}{ input parameters} & \multicolumn{5}{c|}{ meson }
	& \multicolumn{2}{c|}{ $0^+$ diquark }  
	& \multicolumn{1}{c}{ $1^-$ diquark }  \\ 
\hline
$\omega$ $[{\rm GeV}]$ & $D$ $[{\rm GeV}^2]$ & $m_\pi$ & $m_K$ 
                               & $m_{\rho,\omega}$ & $m_{K^*}$     
                               & & $m_{ud}$ & $m_{qs}$ & $m_{ud}$ \\
\hline 
0.40 & 0.93  & 0.138 & 0.495 & 0.742 & 0.936 & & 0.821 & 1.10 & 1.02 \\
0.50 & 0.79  & 0.138 & 0.495 & 0.74  & 0.94  & & 0.688 & 0.96 & 0.89 
\end{tabular}
\caption{\label{tab:masses} Masses (in GeV) of the light mesons
\cite{Maris:1999nt} and diquarks~\cite{Maris:2002yu,Maris:2004bp}
for two different sets of parameters in the effective interaction.}
\end{table}

For electromagnetic interactions, we need the quark-photon vertex, in
addition to the dressed quark propagator and the BSAs.  In impulse
approximation (which we use here), current conservation is guaranteed
as long as this vertex satisfies the vector Ward--Takahashi
identity~\cite{Roberts:1994hh}.  The solution of the inhomogeneous
ladder BSE for the quark-photon vertex satisfies this constraint.
Furthermore, this approach unambiguously includes effects from
intermediate vector mesons, since they appear as poles in the dressed
quark-photon vertex~\cite{Maris:emf}.

\begin{figure}[b]
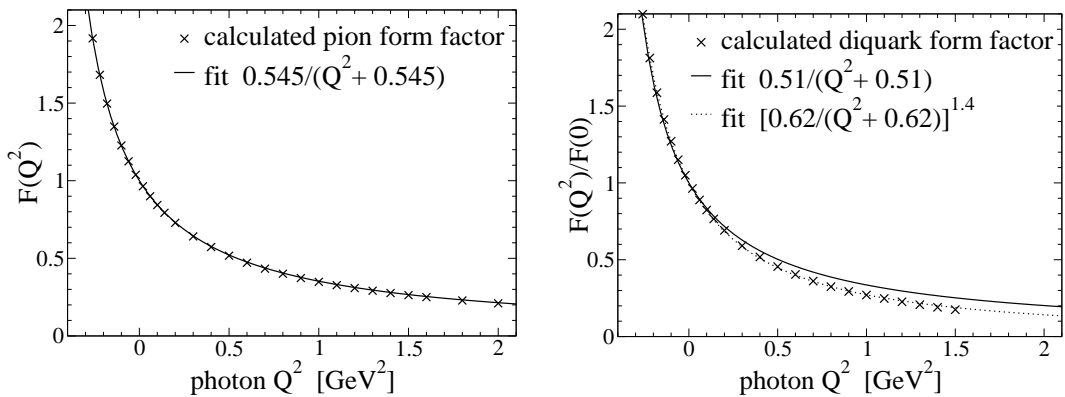

\includegraphics[height=.35\textwidth]{emfpion.eps} \quad
\includegraphics[height=.35\textwidth]{emfuu.eps}
\caption{\label{fig:emfresults} 
The pion (left) and $ud$ diquark (right) form factor, together with
analytic forms fitted in the range $-0.3~{\rm GeV}^2 < Q^2 < 1.5~{\rm
GeV}^2$. (Figure adapted from Ref.~\cite{Maris:2004bp}.)}
\end{figure}

With the quark-photon vertex, we can now calculate the meson and
diquark electromagmetic form factors by coupling the photon to each of
the two (anti-)quarks in the bound state.  The resulting pion form
factor is quite well approximated by a monopole~\cite{Maris:emf}, at
least up to about 2 or 3 GeV.  However, the $ud$ scalar diquark form
factor falls off significantly faster than a monopole, at least at
moderately small spacelike values of $Q^2$, see
Fig.~\ref{fig:emfresults}.  We can obtain a good fit to our
calculation with a form $[M^2/(Q^2 + M^2)]^{1.4}$.  Asymptotically
however, both the pseudoscalar meson and scalar diquark form factors
vanish like $1/Q^2$.

The $ud$ diquark charge radius, $r_{ud} = 0.71~{\rm fm}$, is about 8\%
larger than $r_\pi = 0.66~{\rm fm}$.  This suggests that these
diquarks are somewhat larger in size than pions.  These results appear
to be insensitive to the details of the effective quark-quark
interaction kernel, even though the actual diquark masses are
sensitive to details of the interaction~\cite{Maris:2004bp}.

%
In addition to ordinary mesons and baryons (see e.g. Ref~\cite{review}
and references therein), there could be non-standard hadrons such as
tetraquarks and pentaquarks.  Two color-antitriplet diquarks could
form a color-triplet, which can be combined with an antiquark into a
color-singlet pentaquark~\cite{Jaffe:2003sg}.  Binding between two
diquarks can come from two-quark exchange and meson exchange, but it
is more likely dominated by gluon exchange.  Gluon exchange also
contributes to binding in tetraquarks, i.e. color-singlet
$(qq)$-$(\overline{qq})$ states.  To get an idea about the amount of
binding provided by gluon exchange, we calculate the scalar and vector
bound states of two scalar colored particles in ladder approximation,
ignoring the compositeness of the diquarks.  For the interaction
between the two scalar colored particles, we use the same model as
that for the effective quark-quark interaction.  The color factors are
$f_C = \frac{4}{3}$ for the color-singlet $(qq)$-$(\overline{qq})$
states and $f_C = \frac{2}{3}$ for the color-triplet $(qq)$-$(qq)$
states, cf. the mesons and diquarks.
\begin{table}[h]
\begin{tabular}{ll|ccccc|cccc}
 & & \multicolumn{5}{c|}{color singlet  SU(3)$_f$ nonet}
 & \multicolumn{2}{c}{SU(3)$_f$ triplet}
 & \multicolumn{2}{c}{SU(3)$_f$ anti-sextet}  \\
$m_{qq}$ & $m_{qq}$ 
 & isospin
 & \multicolumn{2}{c}{scalar, $0^+$}
 & \multicolumn{2}{c|}{vector, $1^-$}
 & \multicolumn{2}{c}{scalar, $0^+$}
 & \multicolumn{2}{c}{vector, $1^-$} \\ 
 & & & $M$ & $E_B$ & $M$ & $E_B$
     & $M$ & $E_B$ & $M$ & $E_B$ \\ \hline
0.821 & 0.821 & 0         & 1.068 & 0.574 & 1.331 & 0.311 
                          &  X    &  X    & 1.601 & 0.041 \\
0.821 & 1.10  & $\frac12$, $\frac12$ & 1.342 & 0.579 & 1.592 & 0.329 
                          & 1.624 & 0.297 & 1.853 & 0.068 \\
1.10  & 1.10  & 0, 1      & 1.597 & 0.603 & 1.841 & 0.359 
                          & 1.888 & 0.323 & 2.099 & 0.101 \\ 
\hline
0.688 & 0.688 & 0         & 1.034 & 0.342 & 1.377 &  0.0   
                          &   X   & X     & ---   & ---  \\
0.688 & 0.96  & $\frac12$, $\frac12$ & 1.290 & 0.358 & 1.615 & 0.033 
                          & 1.528 & 0.120 & ---   & ---  \\
0.96  & 0.96  & 0, 1      & 1.530 & 0.390 & 1.844 & 0.076 
                          & 1.770 &  0.150 & ---   & ---  
\end{tabular}
\caption{\label{tab:tetrapenta} Bound state masses and binding
energies in GeV of two scalar constituents in ladder approximation.
The X's indicate that these states are not allowed by the Bose
statistics of the diquarks (the color-triplet states are
anti-symmetric in color indices).}
\end{table}

Our results for the masses and binding energies are given in
Table~\ref{tab:tetrapenta}.  These results suggest that simple
gluon-exchange does not provide enough binding for pentaquarks
dominated by a $(qq)$-$(qq)$-$(\overline{q})$ configuration.
Furthermore, if diquarks are important for pentaquarks, we would also
expect additional states in the meson sector, in particular a scalar
nonet.  Identification of these additional meson-like states however
will be difficult, since they mix with meson molecules and ordinary
mesons.

\begin{theacknowledgments}
%
Part of the computations were performed on the National Science
Foundation Terascale Computing System at the Pittsburgh Supercomputing
Center.
\end{theacknowledgments}


\bibliographystyle{aipproc}   

\bibliography{sample}


\end{document}